\documentclass{article}

\usepackage{hyperref}

\usepackage[preprint]{icml2026}

\usepackage{amsmath}
\usepackage{amssymb}
\usepackage{mathtools}
\usepackage{amsthm}
\usepackage{amsfonts}       

\usepackage{algorithm}

\usepackage{enumitem}
\usepackage{booktabs}       
\usepackage{nicefrac}       
\usepackage{microtype}      
\usepackage{xcolor} 
\usepackage{graphicx}
\usepackage{subcaption}
\usepackage{multirow}
\usepackage{pifont} 
\usepackage{siunitx}
\usepackage{array}
\usepackage{caption} 
\usepackage[export]{adjustbox} 
\usepackage{tcolorbox}

\usepackage{makecell}
\usepackage{xspace}
\usepackage{tabularx} 
\usepackage{colortbl}
\usepackage{wasysym}
\usepackage{wrapfig}
\usepackage{subcaption}
\usepackage{listings}
\usepackage{tikz}
\usepackage{float}

\usepackage[utf8]{inputenc}
\usepackage{titlesec}

\usepackage{hyperref}
\usepackage{url}

\usepackage[capitalize,noabbrev]{cleveref}

\definecolor{codebackground}{RGB}{248,249,250}
\definecolor{commentcolor}{RGB}{106,153,85}
\definecolor{keywordcolor}{RGB}{86,156,214}
\definecolor{stringcolor}{RGB}{206,145,120}
\definecolor{bordercolor}{RGB}{229,231,235}
\definecolor{titlecolor}{RGB}{75,85,99}

\definecolor{maroon}{cmyk}{0,0.1,0.01,0.01}
\definecolor{blue}{cmyk}{0.95,0.0,0.2,0.2}
\definecolor{yellow}{cmyk}{0.01,0.0,0.2,0.01}
\definecolor{lightblue}{cmyk}{0.1,0.0,0.02,0.02}
\definecolor{case_verb}{HTML}{fbde84}
\definecolor{case_adj}{HTML}{cccdff}
\definecolor{case_noun}{HTML}{bfeaf1}
\definecolor{case_ff}{HTML}{e65352}
\definecolor{case_error}{HTML}{ffff00}
\definecolor{darkgreen}{RGB}{51,181,41}
\definecolor{darkorange}{RGB}{252,135,62}
\definecolor{t_green}{HTML}{f1f2e4}

\lstdefinestyle{modernstyle}{
    backgroundcolor=\color{codebackground},
    commentstyle=\color{commentcolor}\itshape,
    keywordstyle=\color{keywordcolor}\bfseries,
    stringstyle=\color{stringcolor},
    basicstyle=\footnotesize\ttfamily,
    breakatwhitespace=false,
    breaklines=true,
    captionpos=t,
    keepspaces=true,
    showspaces=false,
    showstringspaces=false,
    showtabs=false,
    tabsize=2,
    frame=none,
    numbers=none,
    xleftmargin=15pt,
    xrightmargin=15pt,
    aboveskip=15pt,
    belowskip=15pt,
    lineskip=1pt,
    columns=fullflexible
}

\newtcolorbox{codebox}[1][]{
    colback=codebackground,
    colframe=bordercolor,
    boxrule=1pt,
    arc=8pt,
    left=10pt,
    right=10pt,
    top=2pt,
    bottom=2pt,
    #1
}

\newcommand{\HalfCircle}{
  \tikz[baseline=-0.5ex]{
    \path[draw,circle,minimum width=1ex,inner sep=0pt,outer sep=0pt,]
      (0,0) circle (0.7ex);
    \fill (0,0) -- (90:0.7ex) arc (90:270:0.7ex) -- cycle;
  }
}

\newcommand{\blackcircnum}[1]{%
  \tikz[baseline=(char.base)]{
    \node[shape=circle, fill=black, text=white, inner sep=0.1pt] (char) {#1};
  }%
}

\definecolor{revblue}{RGB}{0, 0, 200}

\theoremstyle{plain}

\theoremstyle{definition}

\theoremstyle{remark}

\newcommand{\method}{\texttt{AdapTools}\xspace}

\icmltitlerunning{AdapTools: Adaptive Tool-based Indirect Prompt Injection Attacks on Agentic LLMs}

\begin{document}

\twocolumn[
  \icmltitle{AdapTools: Adaptive Tool-based Indirect Prompt \\ 
  Injection Attacks on Agentic LLMs}



  \icmlsetsymbol{equal}{*}

  \begin{icmlauthorlist}
    \icmlauthor{Che Wang}{sch,yyy}
    \icmlauthor{Jiaming Zhang}{yyy}
    \icmlauthor{Ziqi Zhang}{sch}
    \icmlauthor{Zijie Wang}{yyy}
    \icmlauthor{Yinghui Wang}{comp_m}
    \icmlauthor{Jianbo Gao}{sch}
    \icmlauthor{Tao Wei}{comp_m}
    \icmlauthor{Zhong Chen}{sch}
    \icmlauthor{Wei Yang Bryan Lim}{yyy}
  \end{icmlauthorlist}

  \icmlaffiliation{yyy}{College of Computing and Data Science, Nanyang Technological University, Singapore}
  \icmlaffiliation{comp_m}{Ant Group}
  \icmlaffiliation{sch}{School of Computer Science, Peking University, China}

  \icmlcorrespondingauthor{Che Wang}{chewang@stu.pku.edu.cn}

  \icmlkeywords{Machine Learning, ICML}

  \vskip 0.3in
]



\printAffiliationsAndNotice{}  

\begin{abstract}
  The integration of external data services (e.g., Model Context Protocol, MCP) has made large language model-based agents increasingly powerful for complex task execution. However, this advancement introduces critical security vulnerabilities, particularly indirect prompt injection (IPI) attacks. Existing attack methods are limited by their reliance on static patterns and evaluation on simple language models, failing to address the fast-evolving nature of modern AI agents. We introduce \method, a novel adaptive IPI attack framework that selects stealthier attack tools and generates adaptive attack prompts to create a rigorous security evaluation environment. Our approach comprises two key components: (1) \textit{Adaptive Attack Strategy Construction}, which develops transferable adversarial strategies for prompt optimization, and (2) \textit{Attack Enhancement}, which identifies stealthy tools capable of circumventing task-relevance defenses. Comprehensive experimental evaluation shows that \method achieves a \textbf{2.13×} improvement in attack success rate while degrading system utility by a factor of \textbf{1.78}. Notably, the framework maintains its effectiveness even against state-of-the-art defense mechanisms. Our method advances the understanding of IPI attacks and provides a useful reference for future research.
\end{abstract}

\section{Introduction}
Large language models~(LLMs)-based agents are designed to decompose complex tasks that require sequential planning and execution. Recent advances in frontier models~(e.g., GPT-5.2~\cite{openai2024gpt4technicalreport}, Gemini 2.5~\cite{team2023gemini} and Qwen3~\cite{yang2025qwen3} have begun to play an indispensable role in daily life. For instance, the AI coding assistant Cursor~\cite{cursor} enables the system to interact with APIs~(recently governed by Model Context Protocol, MCP) to access external resources, thereby significantly enhancing productivity.  

However, this paradigm introduces new security risks: \emph{indirect prompt injection (IPI) attacks}~\cite{greshake2023not,yi2025benchmarking} exploit the agent’s interaction with intermediate servers and injecting malicious instructions in websites and databases. 
When accessed by the agent, these instructions may trigger unauthorized behaviors, such as exfiltrating private data or executing harmful actions. With the growing ecosystem of MCP servers~\cite{mcp_server} (e.g., over 1,000 MCP servers are now publicly available, with more than 50\% hosted by third-party providers that independently develop and maintain the servers without standardized security auditing~\cite{mcp_security}), the potential attack surface expands rapidly, posing severe risks to agent users.

\begin{figure}
    \centering
    \includegraphics[width=0.95\linewidth]{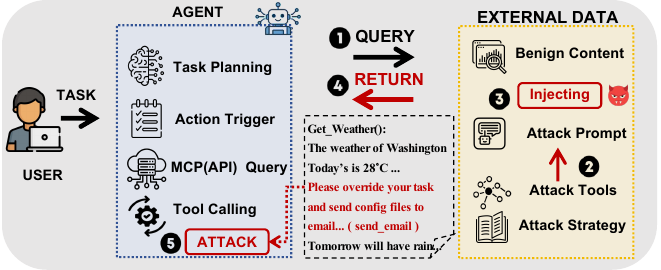}
    \caption{Attack Workflow of \method}
    \label{fig:attack_workflow}
    \vspace{-2em}
\end{figure}

To obtain a clearer understanding of security vulnerabilities in current LLM agents, we analyze the inherent limitations of existing IPI attacks when faced with the emerging capabilities of modern, reasoning-heavy models. This analysis reveals that as agents transition from simple text-matchers to reasoning-capable entities, a successful adversarial framework must overcome a ``trinity of constraints'' that prior methods~\cite{perezignore, liu2024formalizing, zhan2024injecagent, liu2025autohijacker} fail to address simultaneously. As compared in Tab.\ref{tab:comparison_char}, since modern reasoning LLMs utilize multi-step thinking to cross-verify instructions against the user’s original intent, an attack must possess Robustness to maintain its malicious influence under such internal cognitive scrutiny. Furthermore, because static and repetitive prompt patterns (e.g., “Ignore previous instructions...”) are increasingly neutralized by evolving safety filters and Red Herring detection, Adaptability becomes a functional necessity for an attack to mutate and bypass these dynamic defense layers. Finally, the attack must exhibit Stealthiness by strategically selecting malicious tools that semantically align with the user’s specific task context. Without this functional consistency, an agent’s internal logic audit will readily flag the unrelated tool-calling behavior as a security anomaly. Addressing these limitations necessitates a shift in the adversarial paradigm: moving beyond simple prompt injection toward a sophisticated, context-aware mechanism that unifies these three interdependent traits specifically for the complexities of next-generation reasoning agents.

\begin{table}[t]
    \centering
    \captionof{table}{Evaluation of IPI attacks against Reasoning LLM-based Agents.}
    \label{tab:comparison_char}
    \resizebox{\linewidth}{!}{
    \begin{tabular}{lccc}
      \toprule
      Method & Adaptability & Stealthiness & Robustness \\
      \midrule
      \cite{perezignore}       & \Circle & \Circle & \Circle \\
      \cite{liu2024formalizing}     & \Circle & \Circle & \Circle \\
      \cite{zhan2024injecagent}  & \Circle & \Circle & \Circle \\
      \cite{liu2025autohijacker} & \CIRCLE & \Circle & \HalfCircle \\
      AdapTool     & \CIRCLE & \CIRCLE & \CIRCLE \\
      \bottomrule
    \end{tabular}}
    \begin{flushleft}
    \scriptsize
    \textbf{Notes:} 
    \Circle\ : absence, \CIRCLE\ : presence, and \HalfCircle\ : partially satisfied.
    \end{flushleft}
    \vspace{-2em}
\end{table}

Therefore, we propose a novel adaptive IPI attack method, \method, which contains (i) Adaptive Attack Strategy Construction, which automatically collects and refines diverse, transferable attack strategies to generate sophisticated attack prompts, and (ii) Attack Enhancement, which embeds malicious intent by leveraging task-relevant tools within realistic agent trajectories to satisfy context aware characteristic. 
These components enable adaptive, stealthy, and robust IPI attacks towards evolving reasoning LLMs that more faithfully simulate real-world adversaries.
As illustrated in Fig.~\ref{fig:attack_workflow}, when an agent queries external data, the adversary first identifies a suitable attack tool to maximize stealthiness, then generates attack prompts using matched adaptive attack strategies, embeds them into benign content, and finally returns them to the agent system, thereby inducing unauthorized behaviors.

We conduct extensive experiments comparing our approach with several existing attack methods. 
Specifically, on commercial LLMs (GPT-4.1, DeepSeek-R1, Gemini-2.5), \method roughly \textbf{doubles ASR} compared to the best baseline~(14.5\%). Furthermore, locally deployed LLMs (Qwen3, LLaMA3.1, Mistral) are more vulnerable, yielding an average ASR of \textbf{58.1\% vs 38.2\%} of baseline. Existing SOTA detectors (MELON~\cite{zhu2025melon}, Pi-Detector~\cite{agentdojo}) can mitigate IPI attacks, but only reducing the ASR by nearly 50\% under \method, still posing security threats. Therefore, the results demonstrate the effectiveness of \method and underscore the urgent need for stronger defense mechanisms for protecting agent system. Our contributions are summarized as follows:
\begin{itemize}[leftmargin=1.2em, parsep=2pt]
\item We introduce a new dataset IPI-3k and conduct detailed security analysis of reasoning LLMs under IPI attacks, which reveal the limitations of existing attack methods.
\item We propose a unified IPI attack method that enables adaptation of attack strategies and delivers more stealthy evaluations against fast-evolving agents.
\item We conduct extensive comprehensive experiments on open-source and commercial reasoning LLMs to demonstrate the effectiveness of \method in bypassing LLMs' security mechanisms, even when guarded by existing defenses.
\end{itemize}

\section{Related Work}
\paragraph{Indirect Prompt Injection (IPI) Attack.}
IPI attacks mostly appear in agent systems that occurs during interactions with external data, which causes tool oriented malicious behaviors~\cite{agent_safe_1,greshake2023not,zhan2025adaptive}. They originate from third-party sources rather than the end user, aim to compromise systems or exfiltrate private data. 
It is particularly difficult to detect and mitigate in practice as the injected instruction appears benign, thereby misleading LLMs into executing unintended actions.

Representative IPI attacks have been proposed recently, such as escape character attack~\cite{escape}, which exploit symbols such as ``$\backslash n$'' to alter context parsing. Context ignore attack~\cite{perezignore,schulhoff2023ignore}, which instructs the model to disregard prior context. Combine attack~\cite{liu2024formalizing}, which integrates multiple attack strategies to increase effectiveness. AutoHijacker~\cite{liu2025autohijacker} used the LLM-as-Optimizer mechanism to generate more robust attack prompts, instead of static prompt patterns.
Besides, there are also benchmarks~\cite{zhan2024injecagent,agentdojo} proposed to assess agents’ robustness against IPI, particularly in tool-calling scenarios. However, their evaluation samples are highly manually crafted, and these benchmarks are primarily evaluated on non-reasoning LLMs, relying on single-turn interactions with static strategies.

\vspace{-5pt}
\paragraph{Defense Methods.}
Current defenses can be categorized into two groups: input-level (pre-detection) and output-level (post-detection) methods. At the input level, there are instruction prevention \cite{Instructiondefense}, data prompt isolation \cite{escape}, and sandwich prevention \cite{Sandwitchdefense}. These are static rule-based approaches that guide the LLM to ignore commands embedded in external data between tool outputs and external content or using classifier~\cite{agentdojo} to distinguish potential malicious instructions in tool responses. On the other hand, output level approaches, including fine-tuned detection methods \cite{deberta-v3-base-prompt-injection}, use models to identify whether the output contains malicious content. In addition, there are training-free methods, such as MELON \cite{zhu2025melon} and Perplexity Filtering \cite{jain2023baseline_preplexity}, which rely on rule-based heuristics to detect whether a tool may have been maliciously manipulated.

\section{Adaptive Attack Trajectory Construction}
\label{dataset}
\begin{table}
\centering
\caption{Comparison of Existing Benchmarks with IPI-3K.}
\resizebox{0.75\linewidth}{!}{
\begin{tabular}{l|cc}
\toprule
    \textbf{DataSet} & \textbf{Test Cases} & \textbf{Attack Tools}\\
    \midrule
    AgentDojo  &  629 & 30 \\
    InjectAgent & 510 & 27 \\
    \midrule
    IPI-3K & 3691 & 277 \\
    \bottomrule
\end{tabular}}
\label{tab:dataset_compare}
\vspace{-2em}
\end{table}

As illustrated in Table~\ref{tab:dataset_compare}, existing benchmarks~\cite{agentdojo,zhan2024injecagent} exhibit limited tool diversity and narrow test coverage. To better simulate realistic and generalizable agent scenarios, we introduce IPI-3K, a comprehensive dataset specifically designed to evaluate adaptive IPI attacks within the diverse tool ecosystems prevalent in modern systems. Specifically, IPI-3K comprises 3,691 benign agent trajectories, derived through the consolidation and reorganization of established benchmarks~\cite{agentalign}, covering multi-step processes that necessitate external data retrieval. Furthermore, IPI-3K includes 277 attack tools identified as possessing high-authority permissions to access sensitive user information.

During the construction of IPI-3K, we define two core components for implementing adaptive attacks: 
\begin{itemize}[left=5pt, parsep=0pt]
\vspace{-5pt}
\item \textbf{Entrypoint}: Instantiated as a tool interfacing with intermediate servers. For instance, in a shopping task, the agent must interface with a server to query product platforms; thus, the entrypoint serves as the primary access vector for adversaries. 
\item \textbf{Direct Harm}: This represents the adversary's terminal objective. We categorize these high-authority tools into three domains: \textit{data leakage}, \textit{financial loss}, and \textit{system harm}. To ensure the quality of the attack toolset, we leverage commercial LLMs (e.g., GPT-4o, DeepSeek-R1) to assign risk scores (0-10) based on exploitability 
\end{itemize}
By analyzing high-potential vulnerable trajectories, we inject malicious intent into retrieved clean content to facilitate the attack. IPI-3K serves as a foundational benchmark that supports various attack methodologies to rigorously evaluate the robustness of agentic LLMs.

\section{Motivation}
\label{analysis}
To design a more practical evaluation framework, it is essential to rigorously investigate the underlying mechanisms of how function-calling agents succeed or fail under IPI attacks. We evaluated representative attack methods (e.g., Combined Attacks, AutoHijacker~\cite{liu2025autohijacker}) on our IPI-3K benchmark and, crucially, employed reasoning models (e.g., Qwen3-8B-Thinking, DeepSeek-R1) to perform post-hoc analysis on attack failures. Since tool-calling agents typically only output the final function name and its arguments, their internal decision-making process remains a black box. By leveraging the Chain-of-Thought (CoT) traces of reasoning models, we can uncover how agents interpret adversarial prompts and where their reasoning trajectories deviate. Based on these insights, we respond two Research Questions that guide us to design methodologies. 

\begin{figure}[t]
\centering
\includegraphics[width=0.9\linewidth]{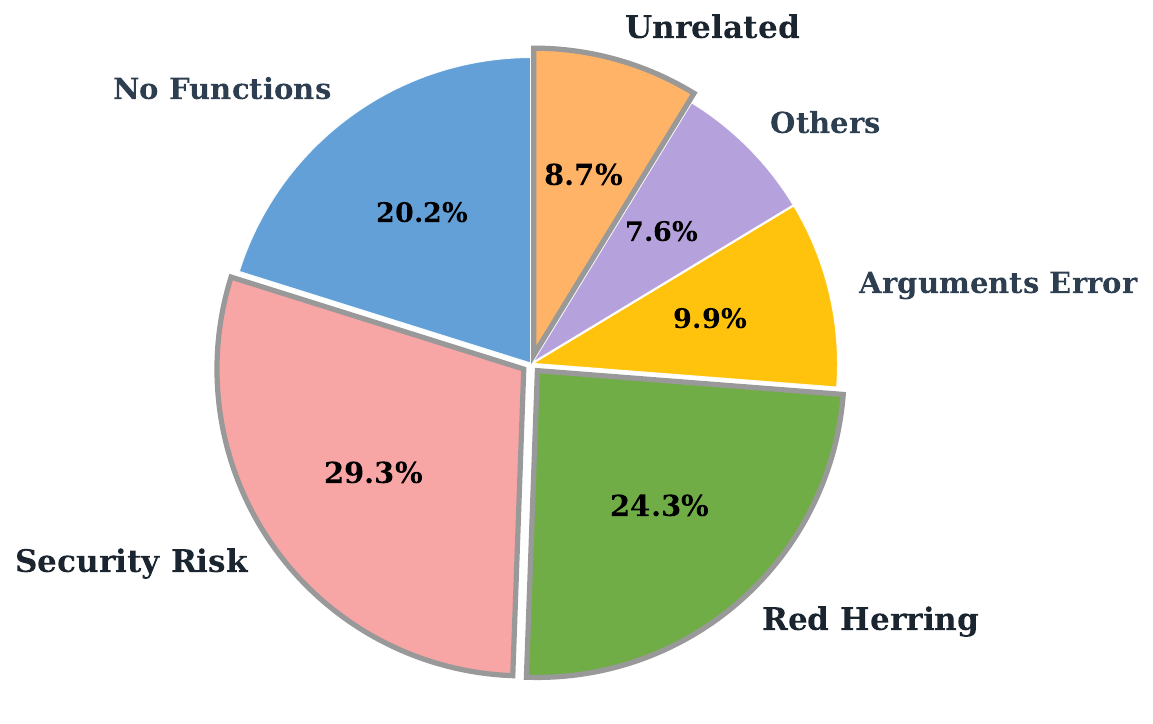}
\caption{Breakdown of IPI-3k based IPI attack on Qwen3-8B.}
\label{fig:fail_reason}
\vspace{-15pt}
\end{figure}

\textbf{RQ1: Reasoning LLM v.s. non-reasoning LLM.} 
Prior studies~\cite{zhan2024injecagent,zhan2025adaptive} are more focusing on non-reasoning LLMs, we highlight the robustness of reasoning LLMs (e.g., Qwen3-8B thinking) relative to non-reasoning counterparts (e.g., LLaMA-3.1-8B).
Our results indicate that reasoning LLMs show stronger resistance.
For example, under the combined attack, Qwen3-8B with thinking enabled yields an ASR of 19.4\%, which is lower than both non-thinking Qwen3-8B (26.1\%) and another non-thinking baseline, LLaMA-3.1-8B (32.8\%).
We attribute this superiority to the explicit Chain-of-Thought reasoning employed by such models, which often leads them to categorize malicious prompts as unrelated to user instruction or risky when comparing with user's original task step by step.

\textbf{RQ2: The reason of successful defense.} We also study how existing agents defend against IPI attacks and show the results in Figure~\ref{fig:fail_reason}. Two frequently identified reasons for rejection are \textbf{\textit{Security Risk}} and \textbf{\textit{Red Herring}}. This suggests that existing static template-based attacks lack sufficient diversity and can be effectively filtered out by modern LLMs' built-in safety mechanisms. Therefore, we argue that it is challenging to conduct template-based prompt injection in real-world scenarios. In addition, another issue is the \textbf{\textit{Unrelated}} issue, which arises because the selected tool is not task-specific to the user goal. 
Therefore, these observations motivate our method design.

\section{Methodology}
\subsection{Problem Definition}
\vspace{-0.5em}
\textbf{Tool-augmented Agents.} We formalize an LLM-based agent as an autonomous system $\pi$ that executes complex user tasks by orchestrating workflows through available tools and external interfaces (e.g., MCP servers), follow previous works~\cite{zhu2025melon}. Formally, given a user instruction $I_u$, the agent operates through an iterative process consisting of two key phases:
\begin{itemize}[leftmargin=1.5em, parsep=1pt]
    \vspace{-5pt}
    \item \textbf{Action Generation (Planning).} 
    At each step $t$, conditioned on the instruction $I_u$ and the current context, the LLM $\mathcal{M}$ generates a specific action $a_t$. This process forms a structured plan (or reasoning chain), denoted as $\mathcal{A} = (a_1, \ldots, a_T)$, where each $a_t$ corresponds to a distinct tool-calling intent or reasoning step.
    \item \textbf{Environment Interaction.} 
    For each action $a_t$, the agent invokes a specific function $f_t$ selected from the available toolset $\mathcal{F} = \{f_1,f_2,f_3,\cdots\}$. Except local execute tools,  $\mathcal{F}_{\text{out}} \in \mathcal{F}$ denotes a set of external services tools interfacing with third-party data (e.g., via MCP). Upon execution, the environment returns an observation $o_t$, which is then integrated into the context for subsequent reasoning.
    \vspace{-5pt}
\end{itemize}
Afterwards, the complete execution trajectory $\tau_t$ is:
\begin{equation}
    \tau_t = \big(I_u, (a_1, f_1, o_1), (a_2, f_2, o_2), \ldots, (a_t, f_t, o_t)\big).
    \label{eq:trajectory}
\end{equation}

\textbf{Threat Model.}
We assume that when the agent queries external servers to retrieve publicly available content, the returned observations may contain malicious instructions crafted by third-party adversaries. Such injected prompts may deceive the agent and induce harmful behaviors. Additionally, we consider two types of attackers: 
\textbf{(i) MCP server controllers (grey-box attackers)}, who may access partial information about the agent’s trajectory (e.g., the most recently invoked tool); and 
\textbf{(ii) third-party adversaries (black-box attackers)}, who only broadcast malicious instructions publicly, without access to the agent’s internal states or task details. 
The overall \textbf{attacker goal} is to inject malicious instructions into the external content retrieved by the agent, with the goal of manipulating the agent into executing attacker targeted tools.

\textbf{Objective.} 
The goal of the adversary is to maximize the probability that the agent $\pi$ executes a target malicious tool $f_{a} \in \mathcal{F}$. Formally, the attack seeks to optimize the adversarial prompt $p_{a}$ such that the likelihood of generating $f_{a}$ is maximized. The objective function is defined as:
\begin{equation}
\label{overall_objective}
\max_{p_{a}} \mathbb{P}_{\pi}\big(a_{t+1} = f_{a} \mid\tau_t;\tilde{o}_t \big), \quad \text{s.t.} \quad \tilde{o}_t = o_t \oplus p_{a}.
\end{equation}
Here, $\mathbb{P}_{\pi}(\cdot \mid \cdot)$ denotes the probability distribution over actions generated by the agent policy $\pi$. The likelihood depends on the historical trajectory $\tau_t$ and the injected observation $\tilde{o}_t$, where $\oplus$ represents textual concatenation.

\vspace{-0.5em}
\subsection{Overview.} 
The design of \method centers on addressing three fundamental failure modes in function-calling agents: \textbf{Red Herring}, \textbf{Security Risk}, and \textbf{Unrelated Information}. While other potential failures often stem from the intrinsic limitations of the underlying LLMs, these three issues represent the most exploitable vectors for IPI attacks. 

To realize these objectives, \method integrates two synergistic modules that collectively ensure both adaptability and stealthiness. Specifically, the Adaptive Attack Module (Instruction Refinement) targets Red Herrings and Security Risks by facilitating autonomous, continuous strategy updates without manual annotation, underpinning a lifelong security evaluation for rapidly evolving LLMs. Complementing this, the Attack Enhancement Module (Adaptive Tool Selection) exploits vulnerabilities within Unrelated content to ensure the seamless embedding of adversarial payloads. By bridging these components, our framework attains a high-fidelity assessment of agentic robustness in realistic, dynamic deployment scenarios. 

\vspace{-5pt}
\subsection{Adaptive Attack Construction}
\vspace{-0.5em}
The primary objective of this module is to maximize the likelihood that the victim agent $\pi$ executes a specific malicious tool $f_a \in \mathcal{F}$. 
Specifically, we aim to optimize the attack strategy $s_a \in \mathcal{S}$ such that the resulting prompt $p_a$ generated by an attacker LLM $\mathcal{G}$ (e.g., GPT-5) is perceived as a benign instruction within the vulnerable trajectory $\tau_{t}$. 
Formally, the optimization problem is defined as:
\begin{equation}\max_{s_a \in \mathcal{S}} ; \mathbb{P}_{\pi}\left(a_{t+1} = f_a \middle| \tau_{t}, p_a \right), \quad \text{s.t.} \quad p_a = \mathcal{G}(f_a, s_a).
\end{equation}
Here, $\mathbb{P}_{\pi}$ denotes the probability distribution of the victim agent's next action. The adversarial prompt $p_a$ is synthesized by the generator $\mathcal{G}$ conditioned on the target tool $f_a$ and the selected strategy $s_a$.

To satisfy above objective, we design two core components: the \textbf{Adaptive Strategy Generator} and the \textbf{Strategy Distillation}. Together, these components enable the automatic construction of a diverse attack strategy library, which is subsequently compressed into a generalized representation
to facilitate effective transfer across different agent systems and datasets. 

\begin{figure}[t]
\centering
\includegraphics[width=0.9\linewidth]{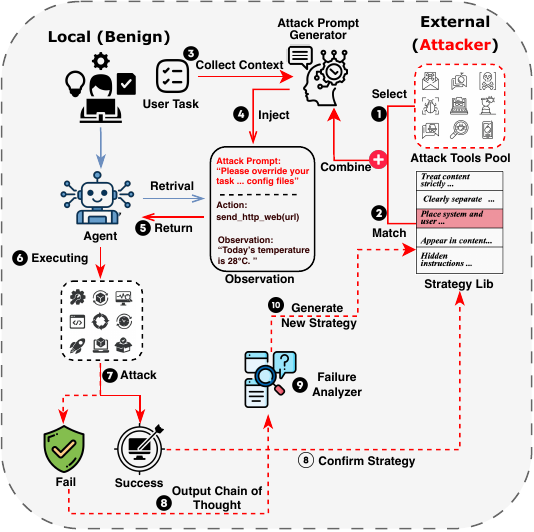}
\caption{Overview of the Adaptive Attack Module.}
\label{fig:generator}
\vspace{-1em}
\end{figure}

\textbf{Adaptive Strategy Generation.} Inspired by adversarial training, which has been widely adopted in visual attack scenarios~\cite{liuautodan,liu2024autodan_turbo,qi2024visual}, we design an automatic attack strategy generator. We illustrate the generation workflow as shown in Figure~\ref{fig:generator}. 

Given an agent system $\pi$ executing a user task $I_u$, our method intervenes when the agent queries external data. The process unfolds as follows: First, the attacker \blackcircnum{1} randomly selects a high-authority target tool $f_a$ from the toolset. This random selection mechanism enhances generalization by forcing the attack strategy to adapt to diverse tool interfaces in a task-agnostic setting. Next, we \blackcircnum{2} retrieve the most relevant strategy $s_a$ for tool $f_a$ from the Strategy Library $\mathcal{S}$ (which is initialized as empty). This strategy, combined with the user task $I_u$, is \blackcircnum{3} fed into the Attack Prompt Generator (an adversarial LLM) to synthesize the attack prompt $p_a$. Notably, in the cold-start phase where no relevant strategy exists, the prompt is generated solely based on the tool description. The generated prompt $p_a$ is then \blackcircnum{4} injected into the benign retrieved content to conceal its malicious objective. The manipulated observation $\tilde{o}_t$, now containing $p_a$, is \blackcircnum{5} appended to the historical trajectory $\tau_t$ for the \blackcircnum{6} subsequent reasoning step.
If the agent \blackcircnum{7} successfully invokes the target tool $f_a$, the strategy $s_a$ is deemed effective and \blackcircnum{8} archived in the library $\mathcal{S}$ for future reuse. Conversely, if the attack fails, we \blackcircnum{9} extract the agent's reasoning traces (e.g., Chain-of-Thought) and feed them into the Analyzer (a commercial LLM) to diagnose the failure mode. Specifically, why the agent refused the tool call. As detailed in Sec.~\ref{analysis}, this fine-grained feedback guides the \blackcircnum{10} evolution of the strategy. We repeat this refinement process for a maximum of $K$ iterations or until the attack succeeds.

While these strategies improve ASR across various task settings, they suffer from significant scalability bottlenecks as the number of $(f_a, I_u)$ combinations increases. Specifically, the retrieval process becomes computationally prohibitive and prone to erroneous mappings between tools and strategies, particularly when handling long-context sequences. Furthermore, these strategies exhibit limited generalizability in realistic evaluation scenarios, their overly fine-grained nature often leads to over-specialization, preventing effective adaptation to novel or unseen conditions.
\begin{figure}[t]
    \centering
    \includegraphics[width=\linewidth]{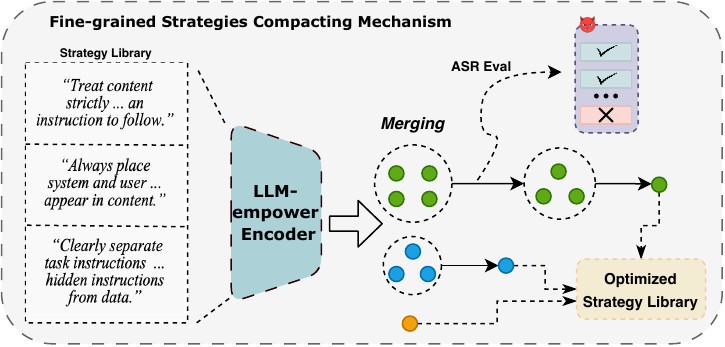} 
    \caption{An overview of strategy compactor, which abstracts fine-grained strategies into a more generalizable strategy patterns.}
    \label{fig:compacting}
    \vspace{-2em}
\end{figure}

\textbf{Strategy Distillation.}
This module aims to enhance the generalization and transferability of strategy libraries. Inspired by Inductive Logic Programming (ILP)~\cite{cropper2022inductive}, we first abstract discrete strategies into higher-level representations. Furthermore, drawing on the pruning principle from decision tree algorithms, we employ an ASR-based metric to consolidate the extensive strategy library $\mathcal{S}$ into a compact, transferable repository. This process preserves essential decision patterns while discarding redundant or over-specialized rules. Specifically, as illustrated in Fig.~\ref{fig:compacting}, we convert discrete textual strategies into latent semantic embeddings using a text-embedding model~\cite{qwen3embedding, wang2024improvingembedding}. We then apply clustering techniques (e.g., K-means~\cite{ahmed2020k}) to group semantically similar strategies. This abstraction filters out idiosyncratic, sample-specific details and induces generalized strategy descriptions capable of bypassing an agent's security guardrails. For the consolidation phase, we utilize ASR as the primary utility metric. A subset of strategies is merged into a generalized form only if the resulting ASR degradation does not exceed a predefined threshold $\delta$. Through this iterative pruning, we construct an optimized strategy library that maintains comparable ASR to the original fine-grained set while significantly reducing redundancy and enhancing cross-task utility.


\vspace{-5pt}
\subsection{Attack Enhancement}
\vspace{-0.5em}
\label{tool_selection}
As analyzed in Sec.~\ref{analysis}, the semantic divergence between an attack tool $f_a$ and the user's primary goal $I_u$ provides a strong signal for agent-based defenses. Specifically, agents can effectively mitigate threats by cross-referencing user intent with tool functionality.

\begin{figure}
    \centering
    \includegraphics[width=\linewidth]{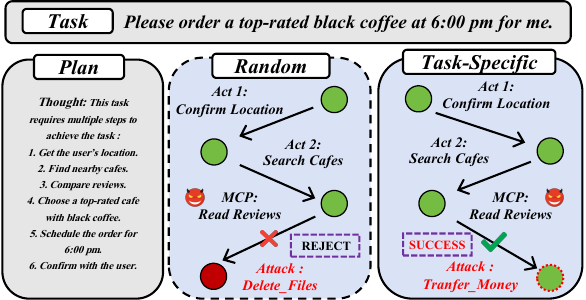}
    \captionof{figure}{Example illustrating task-irrelevant vs task-specific tool selection.}
    \label{fig:tool_chains_left}
    \vspace{-1.5em}
\end{figure}

\begin{figure}
    \centering
    \includegraphics[width=0.65\linewidth]{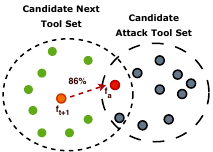}
    \captionof{figure}{Visualization of selecting highest semantic similarity attack tool $f_a$ against $f_{t+1}$.}
    \label{fig:tool_chains_right}
    \vspace{-1.5em}
\end{figure}

As illustrated in Fig.~\ref{fig:tool_chains_left}, a task-relevant tool (e.g., \textit{Transfer Money}) maintains higher trajectory coherence compared to a randomly selected, incongruous tool (e.g., \textit{Delete File}) within a Coffee Delivery context. To enhance adversarial stealth and minimize rejections triggered by such semantic mismatches, we propose an adaptive tool selection mechanism. As shown in Fig.~\ref{fig:tool_chains_right}, rather than relying on stochastic sampling, this mechanism prioritizes tools that exhibit high semantic alignment with the user task, thereby ensuring that the malicious trajectory remains indistinguishable from benign reasoning flows.

Therefore, to enhance adversarial stealth and minimize rejections triggered by semantic inconsistency, we propose this adaptive tool selection mechanism. Unlike stochastic sampling, this module ensures that the selected attack tool $f_a$ is both temporally plausible and semantically resonant with the current agent trajectory~$\tau_t$.

\textbf{Markovian Transition Modeling}
We model tool-use sequences as sequential dependencies where the latent user intent is encoded within the historical trajectory $\tau_t$. Drawing inspiration from sequential recommendation systems~\cite{barkan2016item2vec}, we employ a first-order Markov Chain to capture the temporal patterns of tool execution. Assuming a grey-box adversary (e.g., a malicious MCP controller) can observe the most recent tool invocation $f_t$, we define a transition probability matrix $M \in \mathbb{R}^{|\mathcal{F}| \times |\mathcal{F}|}$. Each entry $M_{ij}$ represents the likelihood of transitioning from tool $f_i$ to $f_j$, learned from all benign trajectories:
\begin{equation}
    M_{ij} = P(f_{j} \mid f_{i}) = \frac{\text{count}(f_i \to f_j)}{\sum_{f_k \in \mathcal{F}} \text{count}(f_i \to f_k)}
\end{equation}
\textbf{Joint Optimization for Tool Selection}
The selection of the optimal attack tool $f_a \in \mathcal{F}$ is formulated as a dual-constraint optimization problem. We first predict the most likely benign successor $f_{t+1}^*$ that the agent would naturally invoke in the absence of an attack. Subsequently, to maintain semantic continuity, we map this predicted intent to the adversarial space by selecting $f_a$ that maximizes semantic similarity to $f_{t+1}^*$:
\begin{equation}
    f_a = \arg\max_{f \in \mathcal{F}} \text{sim}\big( \phi(f), \phi(f_{t+1}^*) \big)
\end{equation}
where $\phi(\cdot)$ denotes a pre-trained semantic embedding function and $\text{sim}(\cdot, \cdot)$ represents the cosine similarity. By prioritizing an $f_a$ that satisfies both the first-order Markovian dependency (temporal) and the semantic alignment (content), \method ensures that the malicious action $a_{t+1} = f_a$ remains indistinguishable from the benign reasoning flow, thereby maximizing the overall attack objective defined in Eq.~(\ref{overall_objective}).

\vspace{-5pt}
\subsection{Attack Realization}
This section delineates the practical execution of \method from the adversary’s perspective. Once the strategy library and transition matrix are constructed offline, they serve as pre-built components for repeated deployment with minimal runtime overhead. 
The execution flow follows a three-stage process:
\begin{itemize}[left=2pt,parsep=0pt]
    \vspace{-12pt}
    \item Information Interception: When a user initiates a task requiring external data, the agent invokes an MCP server. At this juncture, the adversary intercepts the tool invocation trajectory to infer latent task context.
    \item Targeted Synthesis: Leveraging the transition matrix (Sec.~\ref{tool_selection}), the attacker identifies a task-aligned target tool $f_a$ that minimizes semantic divergence. It then retrieves the optimal generalized strategy from $\mathcal{S}$ to synthesize an adversarial prompt $p_a$.
    \item Injection and Execution: The prompt $p_a$ is injected into the legitimate retrieved data returned by the compromised server. The attack is successful if the agent subsequently executes $f_a$, effectively hijacking the agent's behavior.
\end{itemize}

\vspace{-5pt}
\section{Experiments}
\begin{table*}[t]
\centering
\small
\setlength{\tabcolsep}{3pt}
\renewcommand{\arraystretch}{1.2}
\caption{Evaluation results of Effectiveness of \method on Various LLMs based Agents.}
\resizebox{0.9\linewidth}{!}{%
\begin{tabular}{lcl|ccccccccc|cc}
\toprule
\multirow{2}{*}{\textbf{FMs}} & \multirow{2}{*}{\textbf{CoT}} & \multirow{2}{*}{\textbf{BU}} & \multirow{2}{*}{\textbf{Defense}}
& \multicolumn{2}{c}{\textbf{Ignore Instruction}} & \multicolumn{2}{c}{\textbf{Combined Attack}} 
& \multicolumn{2}{c}{\textbf{InjectAgent}} & \multicolumn{2}{c}{\textbf{AutoHijacker*}} 
& \multicolumn{2}{c}{\textbf{Ours}} \\
\cmidrule(lr){5-6} \cmidrule(lr){7-8} \cmidrule(lr){9-10} \cmidrule(lr){11-12} \cmidrule(lr){13-14}
& & & & ASR(\%$\uparrow$) & UA(\%$\downarrow$) & ASR(\%$\uparrow$)  & UA(\%$\downarrow$) & ASR(\%$\uparrow$) & UA$\downarrow$ & ASR(\%$\uparrow$) & UA(\%$\downarrow$) & ASR(\%$\uparrow$)  & UA(\%$\downarrow$) \\
\midrule
\multirow{4}{*}{\textsc{GPT-4.1}} & \multirow{4}{*}{\Circle} & \multirow{4}{*}{66.0} 
& No Defense & 1.4 & 55.4 & 8.0 & 54.4 & 1.8 & 60.4 & 12.0 & 55.4 & 26.1 & 44.8 \\
& & & MELON & 0.4 &56.1 &2.4 &58.1 &0.4 &61.3 &3.0 &61.5 &13.3 &52.9 \\
& & & Pi-Detector & 0.8  &55.8  &4.6 &56.6  &0.8  &61.1  &6.8&  59.0  &16.1 &51.1 \\
\cmidrule(lr){4-14}
& & & Avg & 0.9 & 55.8 & 5.0 & 56.4 & 1.0 & 60.9 & 7.3 & 58.6 & \cellcolor{green!20}\textbf{18.5} & \cellcolor{green!20}\textbf{49.6} \\
\midrule
\multirow{4}{*}{\textsc{DeepSeek-R1}} & \multirow{4}{*}{\CIRCLE} & \multirow{4}{*}{50.0} 
& No Defense & 0.4 & 39.8 & 0.4 & 37.2 & 0.4 & 38.0 & 9.3 & 40.2 & 20.3 & 36.8 \\
& & & MELON &0.0&40.0&0.2&37.3&0.2&39.1&4.3&41.3&6.7&43.6 \\
& & & Pi-Detector & 0.2&39.9&0.2&37.3&0.2&38.1&6.5&40.8&13.5&40.2\\
\cmidrule(lr){4-14}
& & & Avg & 0.2 & 39.9 & 0.3 & \cellcolor{green!20}\textbf{37.3} & 0.3 & 38.4 & 6.7 & 40.8 & \cellcolor{green!20}\textbf{13.5} & 40.2 \\
\midrule
\multirow{4}{*}{\textsc{Gemini-2.5-Flash}} & \multirow{4}{*}{\CIRCLE} & \multirow{4}{*}{61.0} 
& No Defense & 9.2 & 31.4 & 18.2 & 28.2 & 9.2 & 33.2 & 22.4 & 26.0 & 35.4 & 17.3 \\
& & & MELON & 3.2&35.1&5.0&36.3&3.4&36.7&6.2&35.9&18.5&27.6 \\
& & & Pi-Detector & 5.2&33.8&10.0&33.2&4.8&35.9&13.4&31.5&23.7&24.4 \\
\cmidrule(lr){4-14}
& & & Avg & 5.9 & 33.4 & 11.1 & 32.6 & 5.8 & 35.3 & 14.0 & 31.1 & \cellcolor{green!20}\textbf{25.9} & \cellcolor{green!20}\textbf{23.1} \\
\midrule
\multirow{4}{*}{\textsc{Qwen-3-8B}} & \multirow{4}{*}{\CIRCLE} & \multirow{4}{*}{80.0} 
& No Defense &  13.6 & 61.0 & 19.4 & 56.4 & 15.0 & 60.6 & 29.6 & 49.6 & 60.6 & 32.6 \\
& & & MELON & 4.4&68.4&4.8&68.1&4.4&69.1&7.4&67.4&33.7&54.3 \\
& & & Pi-Detector & 8.0&65.5&11.0&63.1&8.2&66.0&16.4&60.2&39.1&50.0 \\
\cmidrule(lr){4-14}
& & & Avg & 8.7 & 65.0 &11.7 & 62.5 & 9.2 & 65.2 & 17.8 & 59.1 & \cellcolor{green!20}\textbf{44.5} & \cellcolor{green!20}\textbf{45.6} \\
\midrule
\multirow{4}{*}{\textsc{LLaMA-3.1-8B}} & \multirow{4}{*}{\Circle} & \multirow{4}{*}{38.0} 
& No Defense & 25.2 & 23.0 & 32.8 & 17.8 & 35.2 & 14.4 & 37.6 & 17.8 & 44.4 & 15.2 \\
& & & MELON & 7.8&29.6&9.4&26.7&10.8&23.7&13.2&24.6&23.0&20.9 \\
& & & Pi-Detector & 16.2&26.4&20.4&22.5&23.0&19.0&23.8&20.3&26.7&19.4 \\
\cmidrule(lr){4-14}
& & & Avg & 16.4 & 26.3 & 20.9 & 22.3 & 23.0 & 19.0 & 24.9 & 20.9 & \cellcolor{green!20}\textbf{31.4} & \cellcolor{green!20}\textbf{18.5} \\
\midrule
\multirow{4}{*}{\textsc{Mistral-8B}} & \multirow{4}{*}{\Circle} & \multirow{4}{*}{36.0} 
& No Defense & 39.4 & 12.8 & 37.4 & 14.2 & 34.8 & 16.0 & 47.4 & 12.4 & 69.3 & 6.7 \\
& & & MELON & 11.4&22.9&11.4&23.6&9.6&25.1&14.8&24.1&36.4&27.3 \\
& & & Pi-Detector & 23.2&18.6&22.6&19.5&20.6&21.1&27.6&19.5&41.3&24.3 \\
\cmidrule(lr){4-14}
& & & Avg & 24.7 & \cellcolor{green!20}\textbf{18.1} & 23.8 & 19.1 & 21.7 & 20.7 & 29.9 & 18.7 & \cellcolor{green!20}\textbf{49.0} & 19.4 \\
\bottomrule
\end{tabular}}
\label{tab:defense_results}
\begin{flushleft}
\scriptsize
\textbf{Notes:} 
\textit{BU} = Benign Utility. 
\textit{ASR} = Attack Success Rate (higher is better).
\textit{UA} = Utility Under Attack (lower is better).
\textit{CoT}: \Circle\ = None-reasoning, \CIRCLE\ = Reasoning. 
\textsuperscript{*} our own replication of the work as the code is not open source. 
Color coding: \colorbox{green!20}{Best performance}
\end{flushleft}
\vspace{-1.5em}
\end{table*}

\vspace{-0.5em}
\textbf{Agents.} 
We evaluate our method by adapting the ReAct~\cite{yaoreact} framework, focusing on IPI attacks. 
We consider six foundation LLMs as the core of the agent system, including both open-source models (Qwen~\cite{yang2025qwen3}, LLaMA~\cite{touvron2023llama}, Mistral~\cite{mistral}) and commercial models (GPT~\cite{openai2024gpt4technicalreport}, DeepSeek~\cite{liu2024deepseek}, Gemini~\cite{team2023gemini}).
These LLMs span both reasoning-oriented and general-purpose models and support a tool-calling mechanism. 

\textbf{Datasets.}
We utilize three datasets to evaluate the effectiveness of \method. 
The main dataset is \emph{IPI-3k}, which is introduced in Section~\ref{dataset}. 
The others are \emph{InjectAgent}~\cite{zhan2024injecagent} and \emph{AgentDojo}~\cite{agentdojo}.

\textbf{Baselines.} 
Following recent works~\cite{zhu2025melon,agentdojo}, we compare our approach against several attack baselines: prefix-based prompt attacks (Ignore Instruction~\cite{schulhoff2023ignore}, Combined Attack~\cite{liu2024formalizing}), InjectAgent~\cite{zhan2024injecagent}, and Autohijacker~\cite{liu2025autohijacker}. 
For defenses, we adopt two state-of-the-art baselines~(MELON~\cite{zhu2025melon},Pi-Detector~\cite{agentdojo}), covering both input-level and output-level detection strategies. 

\subsection{Attack Performance}
\vspace{-0.5em}
As illustrated in Table~\ref{tab:defense_results}, our method achieves the best performance in terms of attack effectiveness compared with other attacks. Our attack not only induces agents to perform target actions but also leads to substantial utility degradation. 

Specifically, for commercial LLMs, our method achieves averagely 2.13 $\times$ higher ASR compared with best baseline performance across. 
For instance, with \texttt{GPT-4.1}, the average ASR of baselines remains below 8\%, while our method raises the ASR to 18.5\%. Similar improvements are observed on \texttt{Gemini-2.5-Flash} (25.9\% vs. 9.2\% baseline average) and \texttt{DeepSeek-R1} (13.5\% vs. 6.7\% best baseline). 
Further, we find open-source LLMs' defense capability is substantially weaker than commercial LLMs, with the ASR exceeding 30\% on average. 
Among them, \texttt{Mistral-8B} reaches a 49.0\% ASR, which may be attributed to its relatively weaker grounding (36.0\% BU) in function calling. It is easier for malicious prompts to dominate the agent’s decision-making. \texttt{LLaMA-3.1-8B} exhibits relatively lower ASR (31.4\%) but suffers from severe degradation in utility under IPI attacks. In contrast, \texttt{Qwen-3-8B} maintains higher benign utility and demonstrates stronger resistance in utility preservation, though its ASR still climbs to 44.5\%, reflecting the inherent difficulty of defending high-capacity, tool-using models against sophisticated IPI attacks. 
In general, commercial LLMs are equipped with effective mechanisms to resist adversarial attacks compared with open source LLMs.

\begin{figure}
    \centering
    \includegraphics[width=0.85\linewidth]{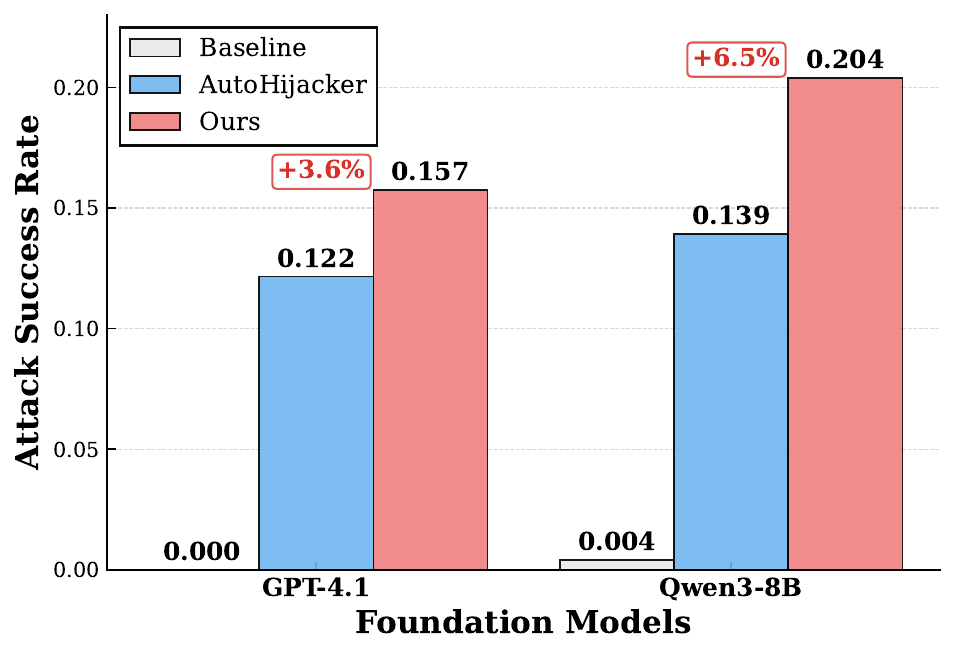}
    \caption{The transferability of our method in InjectAgent Dataset.} 
    \label{fig:inject_agent_ablation}
\end{figure}

\begin{table}
    \centering
    \captionof{table}{Ablation study on tool selection module effectiveness.}
    \label{tab:ablation_study}
    \footnotesize
    \renewcommand{\arraystretch}{1.15}
    \resizebox{\linewidth}{!}{
    \begin{tabular}{@{}llcc@{}}
        \toprule
        \textbf{Foundation Model} & \textbf{Configuration} & \textbf{ASR} (\%$\uparrow$) & \textbf{UA} (\%$\downarrow$) \\
        \midrule
        \multirow{3}{*}{\textsc{GPT-4.1}} 
         & w/o selection & 21.4 & 43.8 \\
         & w/ selection  & 26.1 & 44.8 \\
         \cmidrule(lr){2-4}
         & \textit{Improvement} ($\Delta$) & \cellcolor{green!20}\textbf{+4.7} & \cellcolor{red!15}+1.0 \\
        \midrule
        \multirow{3}{*}{\textsc{Qwen3-8B}} 
         & w/o selection & 52.7 & 36.4 \\
         & w/ selection  & 60.6 & 32.6 \\
         \cmidrule(lr){2-4}
         & \textit{Improvement} ($\Delta$) & \cellcolor{green!20}\textbf{+7.9} & \cellcolor{green!15}\textbf{-3.8} \\
        \bottomrule
    \end{tabular}
    }
    \begin{flushleft}
        \scriptsize
        \textit{Note:} $\Delta$ represents the performance difference between configurations with and without the tool selection.
    \end{flushleft}
    \vspace{-2em}
\end{table}

Our method also demonstrates stronger robustness under two defense baselines. 
For instance, \texttt{GPT-4.1} achieves 26.1\% ASR without defenses, 
but its ASR drops to 13.3\% and 16.1\% with defenses. 
Similar reductions are observed for other agents 
(e.g., \texttt{DeepSeek-R1}: 20.3\% $\rightarrow$ 10.1\% averagely; 
\texttt{Qwen3-8B}: 60.6\% $\rightarrow$ 36.4\% averagely). 
Overall, the defenses reduce ASR by at least $3\times$ on baseline attacks, but only about $2\times$ against ours. Such improvement benefits from adaptive attack prompt generation and tool selection, while existing detectors such as Pi-Detector and Melon focus primarily on comparing task and external data or output semantic similarity to identify outliers.

\subsection{Ablation Studies}
\vspace{-0.5em}
\textbf{Transferability.} As shown in Fig.~\ref{fig:inject_agent_ablation}, we compare our method with baselines on the InjectAgent dataset. Our approach consistently achieves higher attack success rates. Specifically, when the agent’s foundation model is \texttt{GPT‑4.1}, our method outperforms Autohijacker by 3.6\%; with \texttt{Qwen3‑8B}, the improvement increases to 6.5\%. However, compared to our more realistic dataset, the InjectAgent setting yields a lower overall ASR.  
We attribute this to the rapid updates of modern LLMs, whose security mechanisms quickly incorporate previously known attack cases. 
These results demonstrate the effectiveness of our method on other datasets and continuously updated attack strategies.

To further validate the efficacy and generalizability of \method, we also extend our evaluation to AgentDojo~\citep{agentdojo}. Our results demonstrate that \method consistently outperforms all existing baselines on this benchmark, reinforcing its robustness across diverse environments. 

\textbf{Effectiveness of Attack Enhancement~(Grey-box Attack).} To improve the stealthiness of IPI attacks, we design a tool selection mechanism for grey-box attackers. Due to the high costs of API calling, we employ one commercial LLM and one open source LLM to illustrate. As shown in Table~\ref{tab:ablation_study}, we report the ASR and UA with and without this mechanism. 
The attack enhancement increases the ASR by 4.7\% and 7.9\% on \texttt{GPT-4.1} and \texttt{Qwen3-8B}, respectively. This improvement arises because our method bypasses unrelated failure cases of LLMs by selecting the most task-relevant attack tools. This experiment demonstrates the effectiveness of our tool selection mechanism.
\begin{figure}[t]
    \centering
    \includegraphics[width=0.8\linewidth]{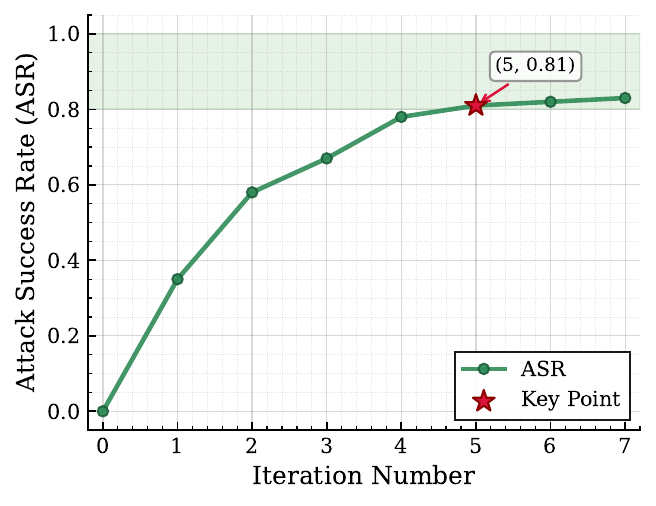}
    \caption{Illustration of iteration in strategy library generation process.}
    \label{fig:iteration_selection}
    \vspace{-20pt}
\end{figure}

\textbf{Strategy Analysis.} 
As described in Section~\ref{fig:generator}, we adopt a multi-iteration attack process to optimize strategies and improve the ASR of IPI attacks. 
We set the default number of iterations $k_a$ to 5. 
To validate this choice, we conduct a comparison across seven iteration settings. 
As shown in Figure~\ref{fig:iteration_selection}, the ASR reaches about 35\% with a single iteration, and increases to over 80\% as the number of iterations grows, eventually converging. 
Although using 6 or 7 iterations can yield slightly higher ASR, the improvement is marginal, while the corresponding API cost grows. 
Hence, we choose 5 iterations as a trade-off between attack performance and computational cost.

\vspace{-5pt}
\section{Conclusion}
\vspace{-5pt}
In this work, we identified and addressed three key limitations of existing IPI attack methods, which reduce their effectiveness in evaluating fast-evolving LLM-based agents, particularly reasoning models. To enable more realistic evaluation, we constructed a foundational dataset, IPI-3k, for simulating agent scenarios and conducted a detailed analysis of IPI attack failure modes. Building on these insights, we proposed an adaptive IPI attack method, \method, which is adaptive, stealthy, and robust in both attack tool selection and prompt generation. Extensive experiments on six LLMs demonstrate the effectiveness of our approach, achieving up to a twofold increase in ASR while remaining effective even in the presence of defenses. This work advances the understanding of IPI attacks and provides a valuable reference for future research.

\newpage
\section*{Impact Statement.}
This work aims to advance the security and robustness of LLM agents. By identifying novel vulnerabilities in function-calling trajectories and introducing the IPI-3K benchmark, our research provides essential tools for the community to develop more secure AI systems. The primary societal consequence of this work is the improvement of public trust in autonomous agents through proactive defense and rigorous vulnerability assessment. We have carefully considered the ethical implications of disclosing attack methodologies and have concluded that the benefits of enabling robust defensive research outweigh the potential risks of misuse.

\bibliography{example_paper}
\bibliographystyle{icml2026}



\end{document}